\documentclass[aps,prx,citeautoscript,superscriptaddress,nofootinbib,eqsecnum,twocolumn,showkeys]{revtex4}  
\usepackage{amsmath,amssymb,bm} 
\usepackage{graphicx}  
\usepackage{color} 
\usepackage[colorlinks=true]{hyperref}  
\hypersetup{
    unicode=false,          
    pdftoolbar=true,        
    pdfmenubar=true,        
    pdffitwindow=false,     
    pdfstartview={FitH},    
    pdftitle={Quantum butterfly effect in weakly interacting diffusive metals},    
    pdfauthor={Aavishkar A. Patel},     
    pdfsubject={},   
    pdfcreator={},   
    pdfproducer={}, 
    pdfkeywords={} {} {}, 
    pdfnewwindow=true,      
    colorlinks=true,       
    linkcolor=magenta, 
    citecolor=blue,        
    filecolor=magenta,      
    urlcolor=blue           
} 

\usepackage{amsfonts}
\usepackage{upgreek}
\usepackage{slashed}
\usepackage{latexsym}
\renewcommand{\vec}[1]{\boldsymbol{#1}}

\def\x{\vec{x}}

\def\k{\vec{k}}
\def\p{\vec{p}}
\def\q{\vec{q}}
\def\Q{\vec{Q}}
\def\R{\vec{R}}
\newcommand{\beq}{\begin{equation}}
\newcommand{\eeq}{\end{equation}}
\def\bea{\begin{eqnarray}}
\def\eea{\end{eqnarray}}

\def \tn {\textnormal}
\newcommand{\nn}{\nonumber \\}

\begin{document}

\title{Quantum butterfly effect in weakly interacting diffusive metals}

\author{Aavishkar A. Patel}
\affiliation{Department of Physics, Harvard University, Cambridge MA 02138, USA}

\author{Debanjan Chowdhury}
\affiliation{Department of Physics, Massachusetts Institute of Technology, Cambridge MA 02139, USA}

\author{Subir Sachdev}
\affiliation{Department of Physics, Harvard University, Cambridge MA 02138, USA}
\affiliation{Perimeter Institute for Theoretical Physics, Waterloo, Ontario, Canada N2L 2Y5}

\author{Brian Swingle}
\affiliation{Department of Physics, Harvard University, Cambridge MA 02138, USA}
\affiliation{Department of Physics, Massachusetts Institute of Technology, Cambridge MA 02139, USA}
\affiliation{Martin Fisher School of Physics, Brandeis University, Waltham MA 02453}

\begin{abstract}
We study scrambling, an avatar of chaos, in a weakly interacting metal in the presence of random potential disorder. It is well known that charge and heat spread via diffusion in such an interacting disordered metal. In contrast, we show within perturbation theory that chaos spreads in a ballistic fashion. The squared anticommutator of the electron field operators inherits a light-cone like growth, arising from an interplay of a  growth (Lyapunov) exponent that scales as the inelastic electron scattering rate and a diffusive piece due to the presence of disorder. In two spatial dimensions, the Lyapunov exponent is universally related at weak coupling to the
sheet resistivity. We are able to define an effective temperature-dependent butterfly velocity, a speed limit for the propagation of quantum information, that is much slower than microscopic velocities such as the Fermi velocity and that is qualitatively similar to that of a quantum critical system with a dynamical critical exponent $z>1$. 
\end{abstract}
\maketitle

\section{Introduction}
\label{intro}

Elucidating the physics of thermalization in isolated quantum systems \cite{Deutsch,Srednicki,Tasaki,Rigol} represents an ongoing challenge in quantum many-body physics, and great progress has been made in recent years due to advances in both theory and experiments \cite{schmied,Blatt14,Greiner,Bloch1,Bloch2,marco1,marco2,monroe1}. In this work we are interested in the process of thermalization in interacting disordered metals, specifically in the physics of quantum information scrambling. Starting from a local perturbation, scrambling describes the spreading of quantum entanglement and information across all of the degrees of freedom in a system \cite{HaydenPreskill,Sekino2008,BrownFawzi,hosur}, leading to a loss of memory of the initial state. The onset of scrambling is associated with the growth of chaos and is an intermediate step in the eventual global thermalization at late times of an isolated quantum many-body system.

It has become clear recently that certain special correlation functions can probe the onset of scrambling \cite{ShenkerStanford2014,kitaevtalk}. While such correlators first appeared in the literature many decades ago \cite{LO69}, there has been a revival in their interest, partly due to their relevance in studying information scrambling in black holes \cite{Shenker2014,ShenkerStanford2014,kitaevtalk}. For two local operators $X$ and $Y$ in a system described by a Hamiltonian $H$, these correlation functions are defined as
\bea
f(t) = \textnormal{Tr}\bigg[\rho~ [X(t), Y]^\dagger~ [X(t), Y]  \bigg],
\eea
where $\rho\propto e^{-H/T}$ is the density matrix of an equilibrium state at temperature $T$ and $X(t)=e^{iHt}Xe^{-iHt}$. The intuition for considering this object is that local operators must grow in time if information is to spread across a system and the commutator measures this growth. Furthermore, in order to access generic matrix elements of the commutator, one considers the average of the square of the commutator, $f(t)$, which is non-negative and avoids phase cancellations. In contrast, the average of the commutator is a response function, and these tend to decay to zero at late times in a chaotic system.

A few comments about $f(t)$ may be helpful. When expanding out $f(t)$ in terms of 4-point functions one finds that it contains both time-ordered and out-of-time-order (OTO) pieces. When dealing with fermionic operators, it is more convenient to study instead the squared anti-commutator. For non-interacting fermions the anti-commutator is proportional to the single particle propagator and encodes causality. More generally, one can relate commutators of composite bosonic operators, e.g. fermion bilinears, to the basic fermion anti-commutator. For a field-theory defined in the continuum, we use the `regulated' version of the correlator above, where two of the operators have been moved halfway along the thermal circle to deal with spurious divergences. 

In a chaotic system with a local Hamiltonian, one expects $f(t)$ to start out small when $X$ and $Y$ are spatially separated, and to grow exponentially in time, $f(t)\sim \epsilon~ e^{\lambda_L t}$, where $\epsilon$ is a small parameter that may depend on time and the distance between $X$ and $Y$. By considering an appropriate analytic continuation of $f(t)$, one can show that there is a fundamental upper bound on $\lambda_L (\leq 2\pi k_B T/\hbar)$ \cite{MSS15}; black-holes and certain random fermion models \cite{SY, kitaevtalk} saturate the bound. On the other hand in glassy systems, or in systems that simply fail to thermalize (but are not fully integrable), $f(t)$ may have a power-law form \cite{BSDC17}. While a measurement of such correlation functions is highly non-trivial, naively requiring a `time-machine' in the laboratory, a few novel protocols have been proposed \cite{BSexp,BSinter,TGexp} and three preliminary experiments \cite{AMR16,Du16,PC16} have already been carried out within the last couple of months.

In this paper, we study scrambling in (weakly) interacting diffusive metals \cite{Lee1985}. We consider the case of Coulomb interactions as well as short-range interactions in two and three spatial dimensions. Based on the general intuition that disorder slows the spread of charge and heat, one might also expect that operators spread more slowly in space in a disordered metal relative to a clean metal. Relatedly, we expect the effects of interactions to be enhanced relative to the clean metal since diffusive electrons move slowly compared to ballistic electrons and the effects of interactions can build up. We compute the growth exponent to lowest order in the strength of the interaction while carrying out an infinite resummation over disorder and indeed find that $\lambda_L$ is larger at low $T$ than the corresponding result for a clean Fermi liquid. We also find that chaos grows in a ballistic fashion, with a velocity that is parametrically smaller than the Fermi-velocity at low temperatures. 

These computations confirm a recent argument by two of us \cite{BSDC17} that even though the transport of charge and energy is diffusive in such metals, generic operators grow ballistically (see also Refs.~\onlinecite{Huse13,Bohrdt2016,Luitz17} for a related observation in one-dimensional systems). This is not too surprising, since there is no reason for the motion of charge and energy to be tied to the growth of chaos in interacting systems; an extreme example being that of a many-body localized (MBL) phase \cite{Huserev,Altmanrev} where partial scrambling occurs even in the absence of any transport of charge and heat \cite{BSDC17,XC16,EFDH16,YC16,HeLu17,Shen16}. When considering long range Coulomb interactions, it is particularly interesting that we find ballistic growth of operators since the microscopic model does not have a Lieb-Robinson bound \cite{LiebRobinson}. There are variants of the Lieb-Robinson bound for systems with power law interactions \cite{Gorshkov14,Monroe14}, but these bounds allow exponential growth of operators with time while we find only linear growth.

\begin{figure}
\begin{center}
\includegraphics[width=0.75\columnwidth]{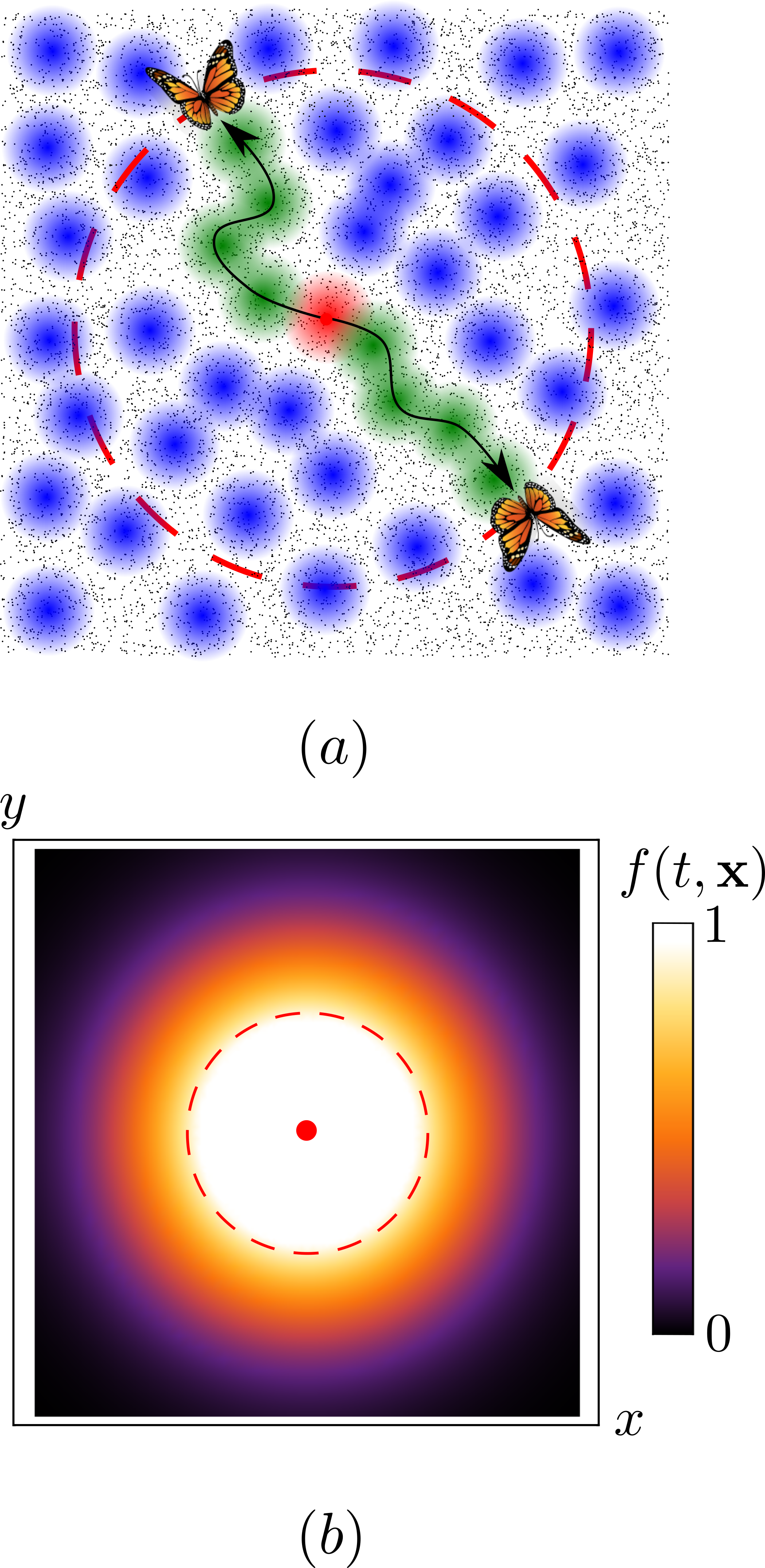}
\end{center}
\caption{(a) Cartoon showing a snapshot at time $t$ of the spread of chaos in an interacting diffusive metal. The fuzzy circles of radius $\propto (Dt)^{1/2}$ represent electrons diffusing through a background of impurities (small black dots). We make an analogy to the spread of an epidemic~\cite{Stanford2016,Huse13}: An `infected' electron inserted into the center of the figure at $t=0$ diffuses outwards (fuzzy red circle). As it encounters other diffusing electrons, it infects them.  These newly infected electrons further infect other electrons and so on (fuzzy green circles). The flight paths of the butterflies track the spread of the infection. The radius of the region containing infected electrons (bounded by the dashed red circle) grows {\it ballistically} as $v_Bt$. Although not shown in the figure, the electrons also have a finite lifespan, given by the inverse of the quasiparticle decay rate. This needs to be taken into account when considering the population of infected electrons as a function of time. The function $f(t,\x)$ is roughly equivalent to the local fraction of infected electrons at a point $\x$. (b) The behavior of $f(t,\x)$ for one operator placed at the center of the figure (red dot) and the other at a position $\x$ shown as a function of $\x$ at a given time $t$. $f(t,\x)$ displays a light-cone (a time slice of which is bounded by the dashed red circle; this region exclusively contains infected particles) within which it has saturated and no longer grows. The radius of this region grows as $v_Bt$.}
\label{introfig}
\end{figure}

On general scaling grounds, the butterfly velocity can be estimated to be $v_B\sim\sqrt{D\gamma_{\mathrm{in}}}$, where $D\sim l^2\tau^{-1}$ is the diffusion constant ($l\equiv$ mean-free path, $\tau^{-1} \approx $ elastic scattering rate) and $\gamma_{\mathrm{in}}$ is a small interaction-induced inelastic scattering rate \cite{BSDC17}. In the presence of weak interactions, we expect 
\begin{equation}
   f(t,\x)\sim e^{\lambda_L t} e^{-{\vec x}^2/(4Dt)}; \label{eq:fdiff}
\end{equation} 
the exponential growth reflects the onset of chaos in an ergodic system as discussed above while the latter contribution is a result of diffusion. Solving for $f(t,\R(t))\sim 1$, where $\R(t)$ is a typical `operator-radius'---which, given an initial perturbation, defines the region in space over which information has spread over time $t$---leads to $\R^2 \sim 4D\lambda_L t^2$ (Figure \ref{introfig}). One therefore obtains a light-cone like growth of $f$ with a butterfly velocity 
\begin{equation}
   v_B = \sqrt{4 D \lambda_L}.
\end{equation}
We show in this paper, by carrying out a perturbative `ladder' computation \cite{Stanford2016}, that the disordered metal does obey Eq.~(\ref{eq:fdiff}),
and the growth exponent, $\lambda_L$ is indeed mostly given by the inelastic scattering rate with a singular temperature dependence. Note that the unitarity of quantum mechanics prevents $f(t,\x)$ from growing to values $\gg 1$ and thus it saturates at very long times. Eq.~(\ref{eq:fdiff}) and the ladder computation are valid only for the pre-saturation growth of $f$.

The rest of this paper is organized as follows: in Section \ref{prelim}, we define our model of interacting electrons in the presence of static disorder and set up the basic elements required for carrying out perturbation theory to leading order in the coupling strength. Section \ref{QC} deals with the perturbative computation of the important terms contributing to $\lambda_L$ and $v_B$ for the case of Coulomb interactions in three spatial dimensions. In Section \ref{ac}, we consider some additional effects in perturbation theory, as well as the case of short-range interactions, and show that our main results are unchanged by these modifications. Finally, in Section \ref{2d}, we study the two-dimensional version of the problem, and point out a subtle difference between $\lambda_L$ and the inelastic scattering rate. Unless explicitly mentioned, $\hbar=k_B=1$ in the rest of this paper.

\section{Preliminaries}
\label{prelim}

We consider a model of $N$ species of electrons in $d \geq 2$ spatial dimensions subject to random potential disorder and weak interactions. We do not take any kind of large-$N$ limit; $N$ is a finite number ($N=2$ for the case of spinful electrons). For most of this work, we shall focus on the physically relevant case of long-range Coulomb interactions in a metal; we also analyze the case of short-range interactions in Section~\ref{ac}. From now on, we focus on the three-dimensional problem with $d=3$ unless otherwise stated, but will analyze the case of two spatial dimensions with $d=2$ in Section~\ref{2d}. 

The Hamiltonian of interest is,
\bea
\label{ham}
H &=& H_0 + H_{\tn{int}},\nn
H_0 &=& \sum_{i=1}^N\int~d^d\x~\psi_i^\dagger(\x)\left(U(\x)-\frac{\nabla^2}{2m}-\mu\right)\psi_i(\x),\nn
H_{\tn{int}} &=& \sum_{i,j=1}^N\int~d^d\x~d^d\x^\prime~V_b(|\x-\x'|) \nn
&~&~~~~~~~~~\times \psi_i^\dagger(\x)\psi_i(\x)\psi_j^\dagger(\x^\prime)\psi_j(\x^\prime),
\eea
where $\psi_i^\dagger(\x)$ ($\psi_i(\x)$) represent fermionic creation (annihilation) operators satisfying the usual anticommutation algebra, $\mu$ is the chemical potential and $m$ is the effective mass of the electrons. The disorder potential $U(\x)$ breaks translational invariance and we assume
\beq
\ll U(\x)U(\x^\prime) \gg = U_0^2~\delta^d(\x-\x^\prime),
\label{disavg}
\eeq
where $\ll ...\gg$ denotes averaging over disorder realizations and $U_0$ denotes the strength of disorder.
We shall treat the interaction, $V_b(|\x-\x'|)$, perturbatively, but will allow for strong disorder via the resummation of various classes of Feynman-diagrams with disorder lines. 
For Coulomb interactions in any number of dimensions $V_b(|\x-\x'|) = e^2/|\x-\x'|$, where $e^2$ will be the small parameter in our perturbative treatment. 

Let us now review the key features of the above theory before setting up the computation for the correlation functions describing chaos in Section \ref{QC}. The remainder of this section closely follows the discussion in standard references (see e.g. Ref.~\onlinecite{Lee1985}).

\begin{figure}
\begin{center}
\includegraphics[width=\columnwidth]{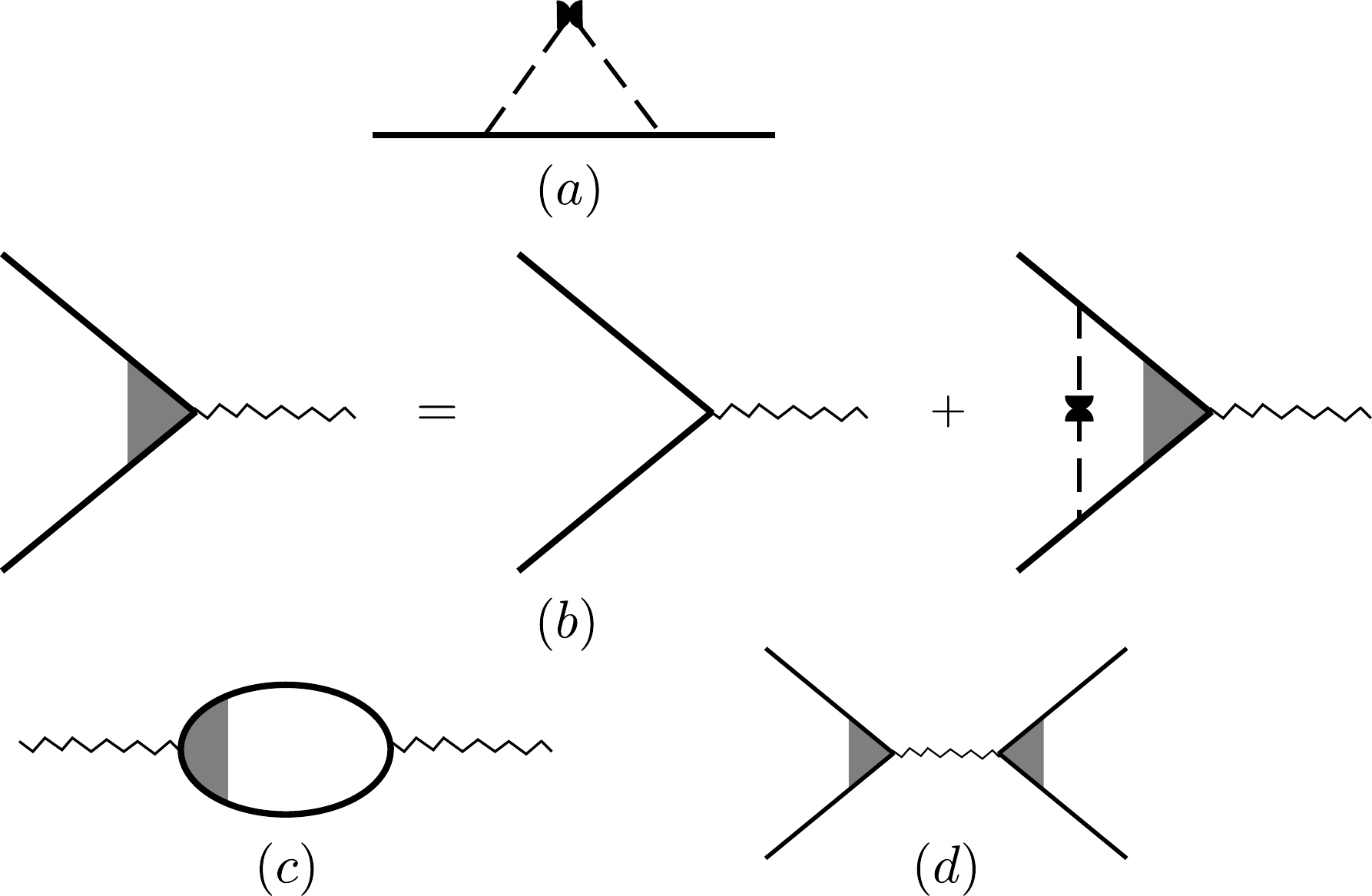}
\end{center}
\caption{(a) The impurity self-energy leading to the elastic lifetime in Eq.~(\ref{G0}) (b) Disorder correction to the electron interaction vertex in Eq.~(\ref{vc}); here, and henceforth the electron lines contain the effect of the impurity self-energy (c) Dynamical screening of the interaction by the disorder-corrected polarization bubble in Eq.~(\ref{screen}) (d) 2-in,2-out process that provides the inelastic electron lifetime; here, and henceforth the interaction line is the dynamically screened interaction.}
\label{basic}
\end{figure}

The bare electron imaginary time Green's function after including the impurity self-energy (Figure \ref{basic}a) is~
\beq
[G_0(\epsilon_n,\p)]^{-1} =-i\epsilon_n + \frac{\p^2}{2m}-\mu-\frac{i}{2\tau}\mathrm{sgn}(\epsilon_n),
\label{G0}
\eeq
where $\tau^{-1} = U_0^2 g(0)$ is the {\it elastic} electron scattering rate due to disorder ($g(0)$ is the density of states at the Fermi level; we use the convention $g(0)=2\pi\int\frac{d^3\p}{(2\pi)^3}\delta(\frac{\p^2}{2m}-\mu)$). 

The real time Green's functions are defined as ($\psi(0)\equiv\psi(0,{\vec{0}})$)
\begin{align}
& \theta(t)\langle\{\psi_i(t,\x),\psi_j^\dagger(0)\}\rangle = i\delta_{ij}G^R(t,\x) \nn
& ~~~~~~~= i\delta_{ij}\int\frac{d^d\k~dk_0}{(2\pi)^{d+1}}G^R(k_0,\k)e^{i(\k\cdot \x-k_0t)}, \nn
&\theta(t)\langle\{\psi_i(t,\x),\psi_j^\dagger(0)\}^\dagger\rangle = -i\delta_{ij}G^{R\ast}(t,\x) \nn 
& ~~~~~~~= -i\delta_{ij}\int\frac{d^d\k~dk_0}{(2\pi)^{d+1}}G^A(k_0,\k)e^{-i(\k\cdot \x-k_0t)}.
\end{align}

As is well known in the theory of non-interacting disordered metals, the disorder averaged product of Green's functions in the particle-hole polarization bubble (density-density correlator) gives rise to the `diffuson' mode at low frequencies and momenta ($|\omega|,v_Fq\ll \tau^{-1}$),
\beq
\Pi(\omega_m,\q) = \frac{dn}{d\mu}\frac{D q^2}{|\omega_m|+D q^2},
\eeq
with the non-interacting diffusion constant, $D \propto l^2/\tau = v_F^2\tau$  ($v_F=\sqrt{2\mu/m}$ is the Fermi velocity). The non-interacting compressibility is $dn/d\mu = Ng(0)/(2\pi)$. In the presence of interactions, the above diffuson mode introduces large vertex corrections (Figure \ref{basic}b) to the electron-interaction vertices
\begin{align}
& \Gamma(\q,\omega_m,\epsilon_n)  = (\theta(\epsilon_n(\epsilon_n-\omega_m)) \nn
&~~ + \theta(\epsilon_n(\omega_m-\epsilon_n))(|\omega_m|+D q^2)^{-1}\tau^{-1}),
\label{vc}
\end{align}
where $\omega_m = 2\pi m T$ and $\epsilon_n = \pi(2n+1) T$ are Matsubara frequencies at a temperature $T$ and effectively screens (Figure \ref{basic}c) the long-range Coulomb interaction to,
\begin{align}
V(\omega_m,\q) &=\frac{4\pi e^2}{q^2}\frac{1}{1+\Pi(\omega_m,\q)\frac{4\pi e^2}{q^2}}\nn
& =\frac{4\pi e^2}{q^2}\frac{|\omega_m|+D q^2}{|\omega_m|+D(K^2 + q^2)}.
\label{screen}
\end{align}
In the above expression, $K^2=4\pi e^2dn/d\mu$ is proportional to the charge compressibility. Note that despite the factor of $e^2$, we still treat $K^2$ as an $\mathcal{O}(1)$ quantity while doing perturbation theory in $e^2$, since $dn/d\mu\propto k_F$ in $d=3$ ($k_F\equiv$ Fermi-momentum) is large. Equivalently, this amounts to doing perturbation theory in $1/(dn/d\mu) \propto 1/(Ng(0))$. Nevertheless we still have $K\ll k_F$.  

Let us also review the computation of the disorder-averaged electron lifetime, which provides the inelastic scattering rate~\cite{Lee1985}. The process in Figure \ref{basic}d, which includes the effects of the dynamically screened interaction and the vertex corrections, gives, via Fermi's Golden Rule, the following expression for the out-relaxation rate or the `inelastic scattering rate' $\gamma_\mathrm{in}(\epsilon)$ for particles with energy $\epsilon$ of a given flavor $i$~\cite{Altshuler1978,Altshuler1979,Efros2012}:   
\begin{align}
\frac{\partial n_{i,\epsilon}}{\partial t}\Bigg|_{\mathrm{out}} & = -Ng(0)\int\frac{d\epsilon^\prime d\Omega}{2\pi^2}\frac{d^3\q}{(2\pi)^3}|V^R(\Omega,\q)|^2 \nn 
& \times \mathrm{Re}\left[\frac{1}{-i\Omega+Dq^2}\right]^2 n_{i,\epsilon} n_F(\epsilon^\prime)(1-n_F(\epsilon-\Omega)) \nn
&~~ \times (1-n_F(\epsilon^\prime+\Omega)) \nonumber \\
&\equiv -n_{i,\epsilon}\gamma_\mathrm{in}(\epsilon).
\label{sig}
\end{align}
Here $n_F(...)$ is the Fermi-Dirac distribution function. Here, the incoming particles are on-shell while the outgoing particles are allowed to be off-shell due to the dynamical interaction $V^R(\Omega,\q)$. At the Fermi level ($\epsilon=0$), this simplifies to
\begin{align}
&\gamma_\mathrm{in}(0) = \frac{Ng(0)}{2\pi}\int_{-\infty}^{\infty}\frac{d\Omega}{\pi}\frac{\Omega}{2\sinh(\beta\Omega)}\int\frac{d^3\q}{(2\pi)^3}|V^R(\Omega,\q)|^2 \nn
&~~~~~~~~~~~~~ \times \mathrm{Re}\left[\frac{1}{-i\Omega+Dq^2}\right]^2 \nn
&\approx \frac{8\pi e^4Ng(0)}{K^4} \int_{-\infty}^{\infty}\frac{d\Omega}{\pi}\frac{\Omega}{2\sinh(\beta\Omega)}\int\frac{d^3\q}{(2\pi)^3}\frac{1}{\Omega^2+D^2q^4} \nn
& \approx \frac{(4-\sqrt{2})\zeta(3/2)e^2T^{3/2}}{4\sqrt{2\pi}d^{3/2}K^2} \approx 0.674\frac{e^2T^{3/2}}{D^{3/2}K^2},
\end{align}
where we made the reasonable assumptions $q\ll K$ and $\Omega\sim T\ll DK^2$. We also used the non-interacting result $Ng(0)\approx 2\pi dn/d\mu$, as corrections due to interactions will only correct $\gamma_\mathrm{in}(0)$ at higher orders in $e^2$. At a finite energy away from the Fermi level, $\gamma_{\mathrm{in}}(\epsilon)\sim\epsilon^{d/2}~h(\epsilon/T)$ in $d-$spatial dimensions, where $h(x)$ is a scaling function of $x$ \cite{Lee1985}.

\section{Many-body quantum chaos}
\label{QC}

To study the onset of quantum chaos for the model introduced in Eq. \eqref{ham}, we compute the flavor-averaged squared anticommutator of electron field operators perturbatively to leading non-trivial order in the coupling $e^2$,
\begin{align}
f(t,\x) = \frac{1}{N}\theta(t) & \sum_{i,j=1}^N  \mathrm{Tr}\left[e^{-\beta H/2}\{\psi_i(t,\x),\psi_j^\dagger(0)\} \right. \nn
& \times \left. e^{-\beta H/2}\{\psi_i(t,\x),\psi_j^\dagger(0)\}^\dagger\right].
\label{oto}
\end{align}
The prefactor of $1/N$ is inserted so that the bare contribution to $f(t,\x)$ is free of factors of $N$. The splitting of $e^{-\beta H}$ into two factors of $e^{-\beta H/2}$ ensures that all operator insertions occur at distinct complex time points, thus avoiding short-distance divergences. The strict positivity of $f(t,\x)$ also guarantees exponential growth at a rate equal to that of the correlator where $e^{-\beta H}$ is not split~\cite{Stanford2016, DCBS17}. These ``regularized" correlators have also been shown to obey fluctuation-dissipation-like relations~\cite{Tsuji2016}. Computing $f(t,\x)$ involves defining the action on a complex-time contour with real time folds separated by $i\beta/2$~\cite{Stanford2016,Aleiner:2016eni,Patel2017,DCBS17}. We must then solve a Bethe-Salpeter equation arising from the resummation of different classes of ladder diagrams to determine $f(\omega,\q)$, which after a Fourier transform yields information about the spatial and temporal structure of growth of chaos. An outline of the derivation of the Feynman rules for Eq.~(\ref{oto}) required to set up the following diagrammatic calculation is presented in Appendix~\ref{FR}.

Let us first quote the results for the non-interacting case, where $V=0$. Here we do not expect chaotic growth of entanglement because the many-body state can be written as a Slater determinant of exact eigenstates of the one-body Hamiltonian. Summing the simplest class of ladder diagrams without any overlapping disorder rungs (Figure~\ref{bs0}) yields the correct qualitative result, as shown in Appendix \ref{NI}. The final answer is
\bea
f(t,\x) \sim f_0(t,\x) + f_1(t,\x) ~e^{-\x^2 / 4Dt},
\eea
where $f_0$ is a rapidly decaying function of time with a rate set by $\tau^{-1}$ and $f_1(t,\x)\sim (Dt)^{-3/2}$ (in $d=3$). At times $t\gg\tau$, $f(t,\x)$ is dominated by the second term, which grows diffusively but then decays as a power law at long times. The diffusive behavior is expected as we have merely computed the particle-hole polarization bubble in real-time. As expected, there is no exponential growth.

We note here an important point, namely that we are actually computing $\ll f(t,\x) \gg$ averaged over different realizations of disorder. In a disordered metal for which the localization length of the eigenstates is far larger than the typical length scale over which the disordered potential varies, the disorder self-averages, and it hence makes sense to consider the disorder average of $f(t,\x)$ within a single copy of a system.
 
\begin{figure}
\begin{center}
\includegraphics[width=\columnwidth]{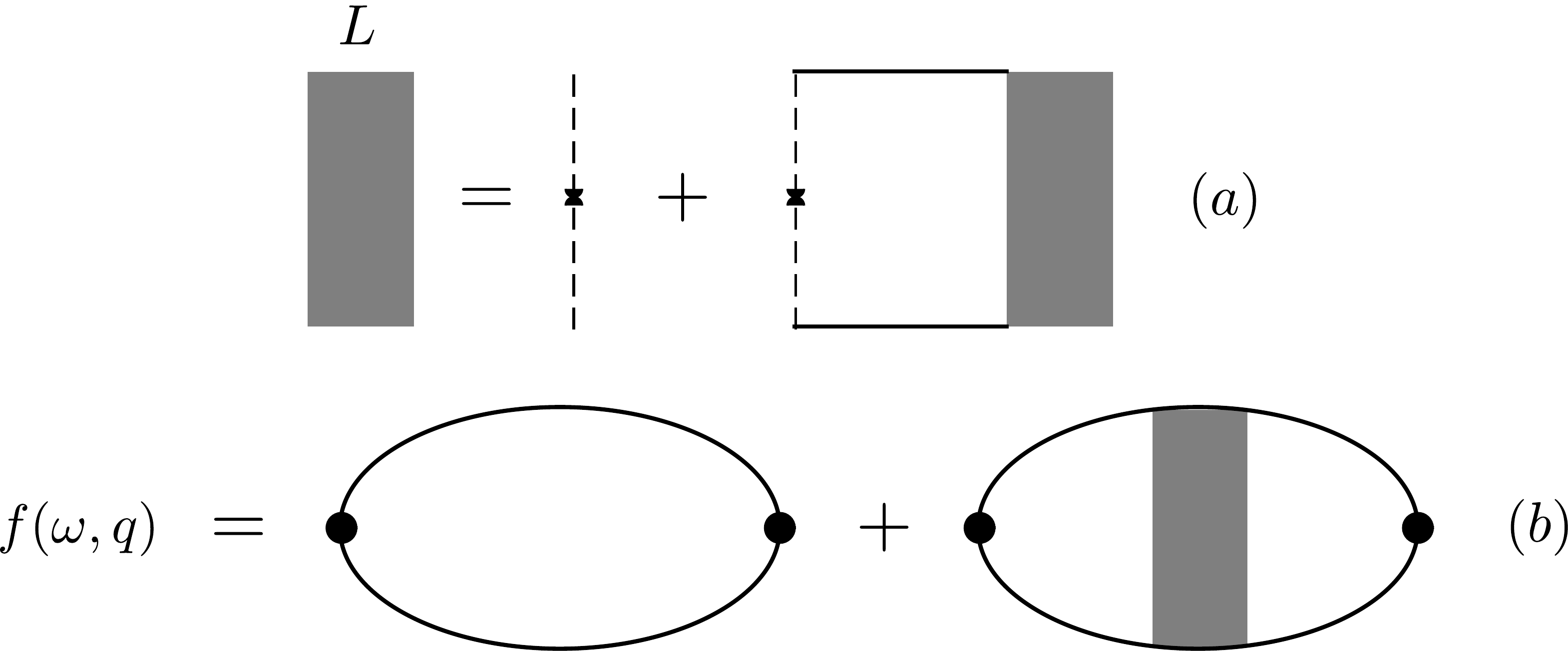}
\end{center}
\caption{(a) Resummation of disorder rungs. (b) Relation between $L(\omega,q)$ and $f(\omega,q)$.}
\label{bs0}
\end{figure}

We now consider the effects of interactions, using a diagrammatic formalism which sums all the singular terms
associated with diffuson and `Cooperon' modes perturbatively in the intereaction strength \cite{Castellani1984,Castellani1986,SS86,SS87}. Our perturbative computation sums all singular disorder corrections while working at $\mathcal{O}(e^2)$ in the interaction, and is formally identical to the theory of Altshuler and Aronov~\cite{Efros2012}. We will examine two effects: (i) dissipative `self-energy' corrections (Figure~\ref{se}) that lead to decay, and, (ii) `ladder' corrections (Figure~\ref{ladder}) that lead to an exponential growth of the squared anticommutator \cite{Stanford2016}. In order to obtain a non-trivial chaotic growth, the effect of the latter has to overwhelm the former. Let us discuss them now one by one.

\subsection{Self-energy corrections}
The non-interacting $L(\omega,\q)$ (Figure \ref{bs0}a) is given by
\beq
L(\omega,\q) =\frac{1}{g(0)\tau^2(-i\omega+Dq^2)}.
\eeq
The dissipative self-energy corrections to the above quantity were considered by Castellani et. al.~\cite{Castellani1984,Castellani1986}. These renormalize $L(\omega,\q)$ at small $\omega,\q$ to
\beq
L(\omega,\q)\rightarrow \frac{1}{g(0)\tau^2}\frac{Z}{-i\omega+\tilde{D}q^2-\Sigma_L^R(0,0)}.
\eeq        
For $T\neq0$, $\Sigma_L(0,0)\neq0$. The field renormalization $Z$ and the renormalization of $D\rightarrow\tilde{D}$ are not of particular concern to us as they will provide a correction to the growth exponent at $\mathcal{O}(e^4)$; from now on we take $Z=1$ and $\tilde{D}=D$. The finite-temperature lifetime {\it is} important, and corrects the growth exponent downwards. 

\begin{figure}
\begin{center}
\includegraphics[width=0.75\columnwidth]{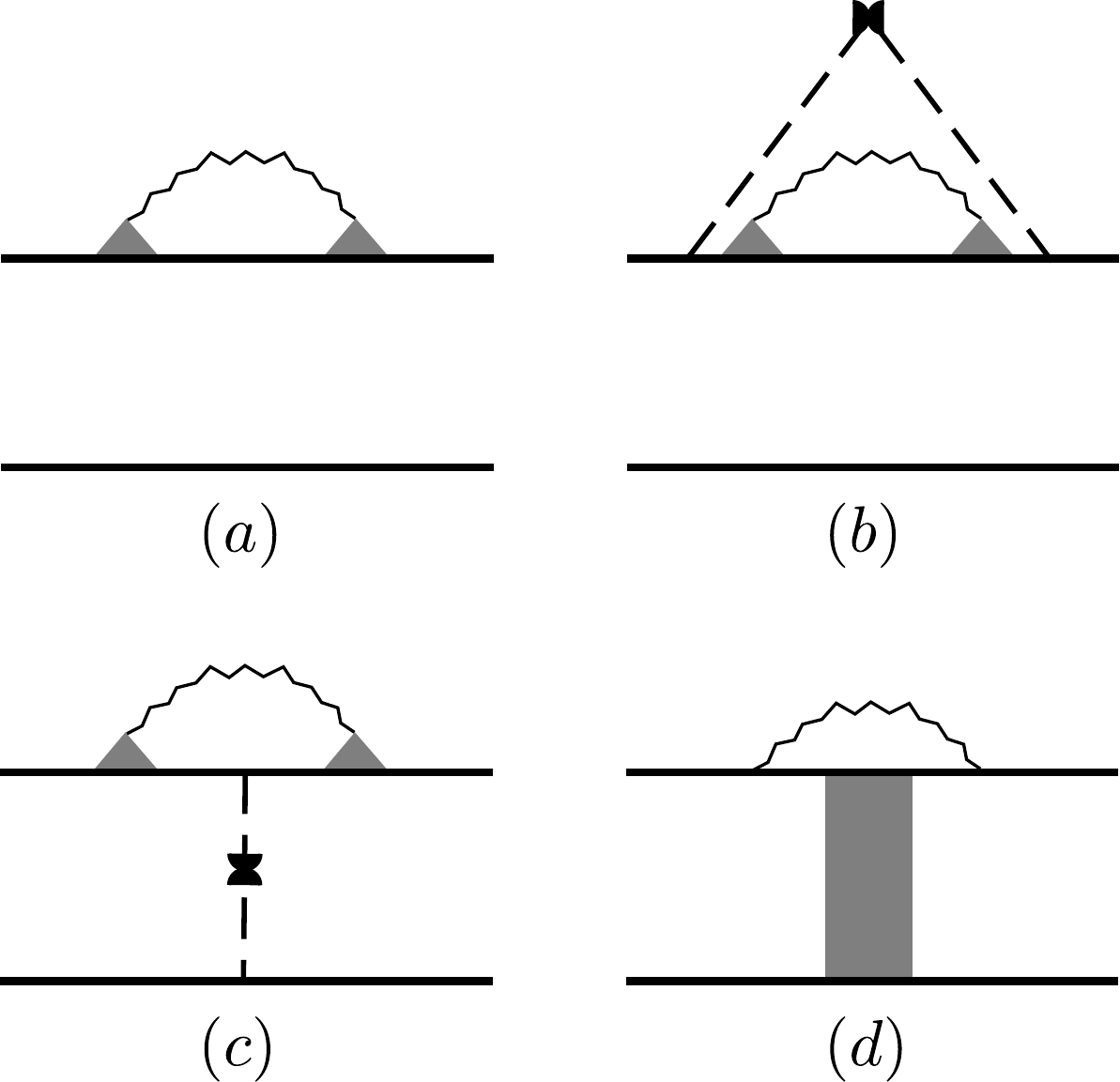}
\end{center}
\caption{The dominant Fock-type self-energy corrections to $L(\omega,\q)$, as described in Refs.~\cite{Castellani1984,Castellani1986}. Each diagram has a partner diagram generated by reflecting about the horizontal axis. Also not shown are the Hartree-type contributions, which are suppressed for sufficiently long-range interactions.}
\label{se}
\end{figure}

To compute $\Sigma_L^R(0,0)$ for the correlator spread across the two time folds, we note the Fock-type diagrams in Figure~\ref{se} (and their partners obtained by reflection about the horizontal axis). We ignore the corresponding Hartree-type diagrams, which are relatively suppressed by a factor of $K^2/k_F^2$~\cite{Efros2012}. In the Fock diagrams, the time folds are connected only by static disorder lines and not the dynamical interaction, and hence there is no distinction between the two time fold correlator and the real-time retarded particle-hole correlator. 
Thus, in the end we only need to focus on the contribution arising from Figure~\ref{se}(d) and its partner, as was noted in Refs.~\cite{Castellani1984,Castellani1986}, to get $\Sigma_L(0,0)$. We have
\begin{align}
& \Sigma_L(\omega_l>0,\q) =  2T\sum_{\Omega_m;~\epsilon_n<\Omega_m<\epsilon_n+\omega_l}\int\frac{d^3\k}{(2\pi)^3}\nn & ~~~~~~~~~~~~~~~~\times \frac{V(\Omega_m,\k)}{D(\k+\q)^2+|\Omega_m+\omega_m|}\Bigg|_{\epsilon_n<0}.
\end{align}
We do this sum by contour integration, noting the branch cut in $V(\Omega_m,\k)$ as $i\Omega_m$ crosses the real axis and that $\epsilon_n$ is a fermionic Matsubara frequency. The non-vanishing contribution upon analytically continuing $\epsilon_n,\omega_l\rightarrow0$ is~\cite{Castellani1986}
\begin{align}
\Sigma_L^R(0,0) & =2g(0)\tau^2\int\frac{d^3\q}{(2\pi)^3}\int_{-\infty}^{\infty}\frac{d\Omega}{\pi}\frac{1}{\sinh\beta \Omega}\nn
& ~~~~~~~~~ \times L(\Omega,\q)\mathrm{Im}[V^R(\Omega,\q)].
\end{align}
We have
\begin{align}
\mathrm{Im}[V^R(\Omega,\q)] & = -\frac{4\pi e^2}{q^2}\frac{DK^2\Omega}{\Omega^2+D^2(K^2+q^2)^2} \nn
& \approx -\frac{4\pi e^2}{q^2}\frac{\Omega}{DK^2}.
\end{align}
Hence
\begin{align}
\Sigma_L^R(0,0) & \approx-\frac{4e^2}{\pi^2DK^2}\int_0^\infty dk\int_{-\infty}^{\infty}dx\frac{1}{\sinh\beta x}\frac{x}{Dk^2-i x} \nn
&=-\frac{2e^2}{\pi D^{3/2}K^2}\int_{-\infty}^{\infty}dx\frac{1}{\sinh\beta x}\frac{x}{\sqrt{-i x}} \nn & =-\frac{(4-\sqrt{2})e^2T^{3/2}\zeta(3/2)}{\sqrt{2\pi}D^{3/2}K^2} \nn 
& \approx -2.695\frac{T^{3/2}}{D^{3/2}K^2}.
\end{align}
Note that $-\Sigma_L^R(0,0)$ is also the decay rate of the Cooperon at zero external pair momentum and frequency~\cite{Fukuyama1983,Castellani1986}, which has been interpreted as the decay rate of electrons in exact eigenstates near the Fermi level~\cite{Abrahams1981,Fukuyama1983,Fukuyama1984}.
\subsection{Ladder diagrams}
\label{LD}
In the ladder diagrams with interaction rungs (Figure~\ref{ladder}), the disorder correction to the interaction vertices occurs on a single time fold. Therefore the second term of Eq. \eqref{vc} does not apply, as it would correspond to the bare interaction vertex connecting Green's functions on opposite time folds before being corrected by disorder, a possibility that is ruled out by the locality of the bare interaction vertex in time. Since the dynamic interaction (which can be interpreted to be mediated by a dynamically fluctuating boson) rung connects two time folds on opposite sides of the thermal circle, its propagator is given by a bosonic Wightman function~\cite{Patel2017,Stanford2016,DCBS17}
\begin{align}
V^W(\Omega,\q) &= \frac{-2\mathrm{Im}[V^R(\Omega,\q)]}{2\sinh\left(\frac{\beta\Omega}{2}\right)} \nn & =\frac{4\pi e^2}{q^2}\frac{D\Omega}{\sinh\left(\frac{\beta\Omega}{2}\right)}\frac{K^2}{\Omega^2+D^2(K^2+q^2)^2}\nn & \approx \frac{4\pi e^2}{q^2}\frac{\Omega}{DK^2\sinh\left(\frac{\beta\Omega}{2}\right)}.
\label{vwight}
\end{align}
Note that only the dynamical part of the interaction $V(\omega_m,\q)-4\pi e^2/q^2$ (which behaves like a Landau-damped boson) contributes to the Wightman function. 

\begin{widetext}
{\it Direct Insertion.-} We first consider the simplest summation of the ladder diagrams with alternating interaction and `diffuson' rungs, $L(\omega,\q)$, given by Figure~\ref{ladder}a. By explicitly considering the series of diagrams, we see that the resulting unit, $F$, depends only upon the frequencies passing through it, but not the momenta. The Bethe-Salpeter equation for $F$ reads
\bea
&& F(\omega,\q,k_0,k_0^\prime) = L(\omega,\q)\delta(k_0-k_0^\prime) + \nn 
&& L(\omega,\q)\int\frac{d^3\k_1d^3\k_2}{(2\pi)^6}\frac{dk_0^{\prime\prime}}{2\pi}~~V^W(k_0-k_0^{\prime\prime},\k_1-\k_2)G_0^R(k_0+\omega,\k_1+\q)G_0^A(k_0,\k_1)\nn
&&~~~~~~~~~~~~~~~~~~~~~~~~~~~~~~~\times G_0^R(k_0^{\prime\prime}+\omega,\k_2+\q)G_0^A(k_0^{\prime\prime},\k_2)F(\omega,\q,k_0^{\prime\prime},k_0^\prime).
\eea
The overall sign of the rung term is $+1$, coming from $i^2(-i)^2$, where the factors of $i$ are generated by the Hubbard-Stratonovich transformation of the Coulomb interaction to a fermion-boson interaction and the two real-time fermion-boson interaction vertices.  After some manipulations, and assuming $\k_1, \k_2$ are close to the Fermi surface, this becomes 
\bea
&& F(\omega,\q,k_0,k_0^\prime) \approx\frac{1}{g(0)\tau^2}\frac{1}{(-i\omega+Dq^2)}\delta(k_0-k_0^\prime) + \nn 
&& \frac{m^2}{g(0)\tau^2}\frac{1}{(-i\omega+Dq^2)}\frac{4\pi e^2}{DK^2}\int\frac{d\epsilon_1d\epsilon_2}{(2\pi)^4}\frac{dk_0^{\prime\prime}}{2\pi}\frac{k_0-k_0^{\prime\prime}}{\sinh\left(\frac{k_0-k_0^{\prime\prime}}{2T}\right)}\ln\left(\frac{16\mu^2}{(\epsilon_1-\epsilon_2)^2}\right)\nn
&&~~~~~~~~~~~~~~~~~~~~~~~~~~~~~~~~~~~~~\times\frac{1}{(\epsilon_1-k_0)^2+\frac{1}{4\tau^2}}\frac{1}{(\epsilon_2-k_0^{\prime\prime})^2+\frac{1}{4\tau^2}}F(\omega,\q,k_0^{\prime\prime},k_0^\prime),
\eea
where we also set $\omega,q=0$ in the internal Fermion Green's functions, because the leading dependence on $\omega,q$ for small $\omega,q$ comes from the $1/(-i\omega+Dq^2)$ multiplying the integral. We can rewrite this for small $k_0,k_0^{\prime\prime}\ll\tau^{-1}\ll\mu$ as 
\beq
F(\omega,\q,k_0,k_0^\prime) \approx\frac{1}{g(0)\tau^2}\frac{1}{(-i\omega+Dq^2)}\delta(k_0-k_0^\prime) + \frac{m^2}{g(0)}\frac{\ln(4\mu\tau)}{(-i\omega+Dq^2)}\frac{2e^2}{\pi DK^2}\int\frac{dk_0^{\prime\prime}}{2\pi}\frac{k_0-k_0^{\prime\prime}}{\sinh\left(\frac{k_0-k_0^{\prime\prime}}{2T}\right)}F(\omega,\q,k_0^{\prime\prime},k_0^\prime).
\label{Fker}
\eeq
As a matrix equation
\beq
\mathbb{F} = \frac{\mathbb{I}/(g(0)\tau^2)}{(-i\omega+Dq^2)\mathbb{I}-\frac{2e^2\ln(4\mu\tau)}{v_FDK^2}\mathbb{A}_0},
\eeq
where the elements of $\mathbb{A}_0$ are given by the integral kernel of the previous equation \eqref{Fker}. Note that the translationally invariant structure of $\mathbb{A}_0$ implies plane wave eigenstates. The growing part of $f(\omega,\q)$ is obtained by appending external lines to $F$, capping off the ladder sum and integrating over momenta (which just provides two factors of $g(0)\tau=\int\frac{d^3\k}{(2\pi)^3}G_0^R(k_0,\k)G_0^A(k_0-\omega,\k)$ for $\omega\ll\tau^{-1}$) and frequencies (Figure~\ref{ladder}c): $f(\omega,\q)=(g(0)\tau)^2 \int\frac{dk_0}{2\pi}\frac{dk_0^{\prime}}{2\pi}F(\omega,\q,k_0,k_0^\prime)$. Therefore $\mathbb{F}$ and $\mathbb{A}_0$ have the same eigenvectors and the largest positive eigenvalue of $\mathbb{A}_0$ ($=\pi T^2$) provides the growth exponent 
\beq
\lambda_L^{(0)} \approx \frac{2\pi e^2}{v_F DK^2}T^2\ln(4\mu\tau).
\eeq
Thus the growth exponent produced by the simplest `direct' ladder insertion considered above is insufficient to overwhelm the $T^{3/2}$ decay rate from the self-energy corrections. We need to thus consider other ladder insertions at $\mathcal{O}(e^2)$ and check to see if they generate an exponent that successfully competes with the decay rate. Henceforth, we ignore the contribution of $\mathbb{A}_0$ to the ladder sum.

\begin{figure}
\begin{center}
\includegraphics[height=2.5in]{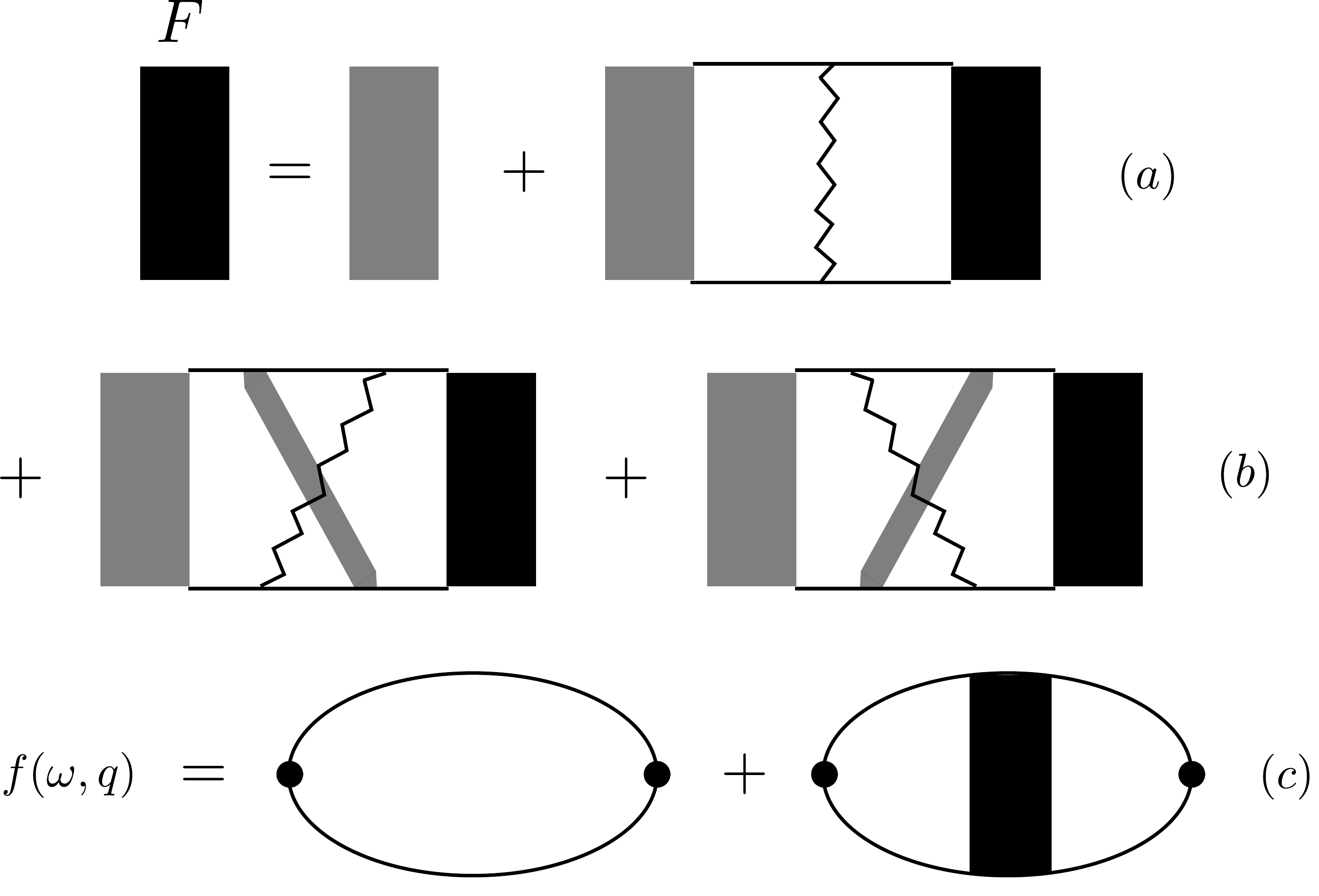}
\end{center}
\caption{Ladder insertions at $\mathcal{O}(e^2)$ which provide exponentially growing contributions to $f(t,\x)$. The `direct' insertion in (a) provides a contribution that grows at a rate proportional to $T^2$, slower than the `exchange' insertions in (b), which grow as $T^{3/2}$. The relationship between the function $f(\omega,\q)$ and the ladder series is shown in (c).}
\label{ladder}
\end{figure}

{\it Exchange insertion.-} As discussed above, we need to consider additional ladder insertions at the same order in perturbation theory which at least compete with the previously computed decay rate. At $\mathcal{O}(e^2)$, these come from Figure~\ref{ladder}b. The sum of the two insertions gives the following integral equation:
\bea
&&F(\omega,\q,k_0,k_0^\prime) = L(\omega,\q) \delta(k_0-k_0^\prime) + \nn
&&L(\omega,\q)\int\frac{d^3\k_1}{(2\pi)^3}\frac{d^3\k_2}{(2\pi)^3}\frac{d^3\k_3}{(2\pi)^3}\frac{dk_0^{\prime\prime}}{2\pi}~V^W(k_0-k_0^{\prime\prime},\k_2-\k_3)L(k_0+\omega-k_0^{\prime\prime},\k_2+\q-\k_3)G^R_0(k_0+\omega,\k_1+\q)G^A(k_0,\k_1) \nn 
&&\times G^R_0(k_0+\omega,\k_2+\q) G^A_0(k_0^{\prime\prime},\k_1-\k_2+\k_3)G^R_0(k_0^{\prime\prime}+\omega,\k_3+\q)G^A_0(k_0^{\prime\prime},\k_3)F(\omega,\q,k_0^{\prime\prime},k_0^{\prime})\nn
&&+L(\omega,\q)\int\frac{d^3\k_1}{(2\pi)^3}\frac{d^3\k_2}{(2\pi)^3}\frac{d^3\k_3}{(2\pi)^3}\frac{dk_0^{\prime\prime}}{2\pi}~V^W(k_0-k_0^{\prime\prime},\k_3-\k_2)L(k_0^{\prime\prime}+\omega-k_0,\k_2+\q-\k_3)G^R_0(k_0+\omega,\k_1+\q)G^A(k_0,\k_1) \nn 
&&\times G^R_0(k_0+\omega,\k_2+\q) G^A_0(k_0^{\prime\prime},\k_1-\k_2+\k_3)G^R_0(k_0^{\prime\prime}+\omega,\k_3+\q)G^A_0(k_0^{\prime\prime},\k_3)F(\omega,\q,k_0^{\prime\prime},k_0^{\prime})
\label{ladder2}.
\eea
The overall sign of this contribution is $+1$ for the same reasons as above. Moreover, the two contributions are equal to each other. As before, we ignore the small $\omega, \q$ contribution coming from within the integrand, and throw out the short-wavelength/high-frequency parts of the interaction. Since the interaction is long-ranged, the largest contribution to the integrals comes when the momentum $\k_2-\k_3$ appearing in the internal interaction and in the `diffuson' rungs is small compared to the momenta flowing through the internal fermion lines, which are $\mathcal{O}(k_F)$. We thus shift $\k_3\rightarrow \k_3+\k_2$ and then ignore $\k_3$ everywhere except in the interaction and `diffuson' rungs, which are singular at small $\k_3$. Then we have,
\bea
&&F(\omega,\q,k_0,k_0^\prime)\approx \frac{1}{g(0)\tau^2}\frac{1}{(-i\omega+Dq^2)}\delta(k_0-k_0^\prime) + \nn
&&\frac{8\pi e^2}{\tau^4(-i\omega+Dq^2)K^2}\int \frac{d^3\k_3}{(2\pi)^3}\frac{d\epsilon_1d\epsilon_2}{(2\pi)^2}\frac{dk_0^{\prime\prime}}{2\pi}\frac{k_0-k_0^{\prime\prime}}{\sinh\left(\frac{k_0-k_0^{\prime\prime}}{2T}\right)}\frac{1}{(\epsilon_1-k_0)^2+\frac{1}{4\tau^2}}\frac{1}{(\epsilon_2-k_0^{\prime\prime})^2+\frac{1}{4\tau^2}}\frac{1}{\epsilon_1-k_0^{\prime\prime}+\frac{i}{2\tau}} \nn
&&\times\frac{1}{\epsilon_2-k_0-\frac{i}{2\tau}}\frac{1}{D^2k_3^4+(k_0-k_0^{\prime\prime})^2}F(\omega,\q,k_0^{\prime\prime},k_0^\prime) \nn
&&\approx \frac{1}{g(0)\tau^2}\frac{1}{(-i\omega+Dq^2)}\delta(k_0-k_0^\prime) + \frac{4e^2}{\pi(-i\omega+Dq^2)K^2}\int_0^\infty dk_3\int_{-\infty}^\infty \frac{dk_0^{\prime\prime}}{2\pi}\frac{k_0-k_0^{\prime\prime}}{\sinh\left(\frac{k_0-k_0^{\prime\prime}}{2T}\right)}\frac{k_3^2}{D^2k_3^4+(k_0-k_0^{\prime\prime})^2}F(\omega,\q,k_0^{\prime\prime},k_0^\prime) \nn
&&\approx \frac{1}{g(0)\tau^2}\frac{1}{(-i\omega+Dq^2)}\delta(k_0-k_0^\prime) + \frac{e^2\sqrt{2}}{(-i\omega+Dq^2)D^{3/2}K^2}\int \frac{dk_0^{\prime\prime}}{2\pi}\frac{k_0-k_0^{\prime\prime}}{\sinh\left(\frac{k_0-k_0^{\prime\prime}}{2T}\right)}\frac{1}{\sqrt{|(k_0-k_0^{\prime\prime})|}}F(\omega,\q,k_0^{\prime\prime},k_0^\prime).
\eea
This gives the matrix equation
\beq
\mathbb{F} = \frac{\mathbb{I}/(g(0)\tau^2)}{(-i\omega+Dq^2)\mathbb{I}-\frac{e^2\sqrt{2}}{D^{3/2}K^2}\mathbb{A}_1},
\eeq
where the matrix elements of $\mathbb{A}_1$ are given by the integral kernel in the last line of the above equation. As was the case with $\mathbb{A}_0$, the largest positive eigenvalue of $\mathbb{A}_1$ comes from an eigenvector with constant entries. We thus obtain the net growth exponent after taking into account the dissipative self-energy:
\beq
\lambda_L^{(1)}=\frac{e^2T^{3/2}(4-\sqrt{2})\zeta(3/2)}{\sqrt{\pi}D^{3/2}K^2}+\Sigma_L^R(0,0) = \frac{e^2T^{3/2}(5-3\sqrt{2})\zeta(3/2)}{\sqrt{\pi}D^{3/2}K^2} \approx 1.116\frac{e^2T^{3/2}}{D^{3/2}K^2}.
\eeq
Hence  
\beq
f(\omega,\q) = (g(0)\tau)^2 \int\frac{dk_0}{2\pi}\frac{dk_0^{\prime}}{2\pi}F(\omega,\q,k_0,k_0^\prime) = \frac{g(0)}{(2\pi)^2}\frac{1}{-i\omega+Dq^2-\lambda_L^{(1)}}.
\eeq
This returns Eq.~(\ref{eq:fdiff}) after a Fourier transform.  

\section{Additional considerations}
\label{ac}
In the previous section, we computed the squared anticommutator and the leading ${\cal{O}}(e^2)$ correction to the growth exponent by doing an infinite resummation of the disorder lines. It is natural to ask the following questions: (i) Do ladder diagrams with a different skeleton structure of the disorder lines affect the exponent? (ii) What is the contribution of the other diagrams at ${\cal{O}}(e^2)$ that have been ignored in Figure \ref{ladder} above? (iii) How sensitive are the above results to the specific form of the (Coulomb) interaction, $V(|{\vec r} - {\vec r}'|)$?

We address all of these concerns one by one in this section.
\subsection{Crossed disorder rungs}

Instead of using the `diffuson' rung, $L(\omega,\q)$, considered thus far, we can sum diagrams with `maximally-crossed' disorder rungs (Figure~\ref{mc}). As is well known, this gives
\beq
L_c(\omega,\Q) = \frac{1}{g(0)\tau^2(-i\omega+DQ^2)},
\eeq
where $\Q$ is the $\it total$ momentum of the incoming or outgoing particle-particle pairs. As with $L(\omega,\q)$, $\omega$ is still the net lateral frequency transfer above as the disorder rungs cannot transfer frequency. At the non-interacting level, this gives
\beq
f_c(\omega,0) = \int\frac{d^3\k}{(2\pi)^3}\frac{d^3\k^\prime}{(2\pi)^3}\frac{dk_0}{2\pi}\frac{1}{g(0)\tau^2(-i\omega+D(\k+\k^\prime)^2)}G^R_0(k_0,\k)G^A_0(k_0-\omega,\k)G^R_0(k_0,\k^\prime)G^A_0(k_0-\omega,\k^\prime).
\eeq
It is easy to see that this expression does not have a pole at small $\omega$, and hence we do not need to consider contributions with $L_c$ as the base unit (In two spatial dimensions, there is a logarithmic singularity at small $\omega$ that is still weaker than the pole in the contribution with $L$). We can also insert $L_c$ as an internal rung in the series with $L$ as the base unit, such as by replacing $L(k_0-k_0^{\prime\prime},\k_2-\k_3)\rightarrow L_c(k_0^{\prime\prime}-k_0,\k_1+\k_3)$ in the integrand of Eq.~(\ref{ladder2}). However, in this case, the same small momentum then does not appear in both the interaction and $L_c$ rungs, and the resulting contribution is thus less singular than the one in Eq.~(\ref{ladder2}), scaling as subleading powers of $T$ starting at $T^2$.

Similarly, we can consider insertions such as those in Figure~\ref{ladder}, but with additional internal $L$ rungs. These are also less singular than the ones shown for the same reason.

\begin{figure}
\begin{center}
\includegraphics[height=1.25in]{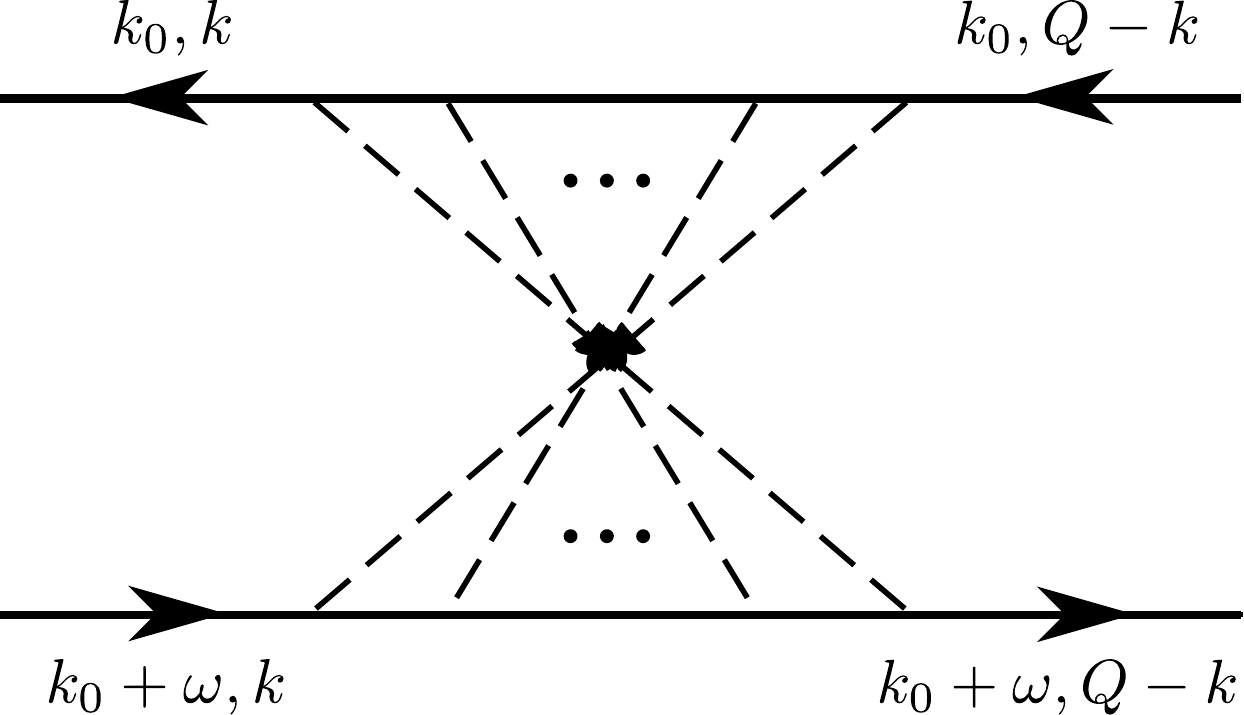}
\end{center}
\caption{A diagram in the `maximally-crossed' series. The sum of this series gives $L_c(\omega,\Q)$ as discussed in the main text.}
\label{mc}
\end{figure}

\subsection{Additional diagrams at ${\cal{O}}(e^2)$}

\begin{figure}
\begin{center}
\includegraphics[height=2.5in]{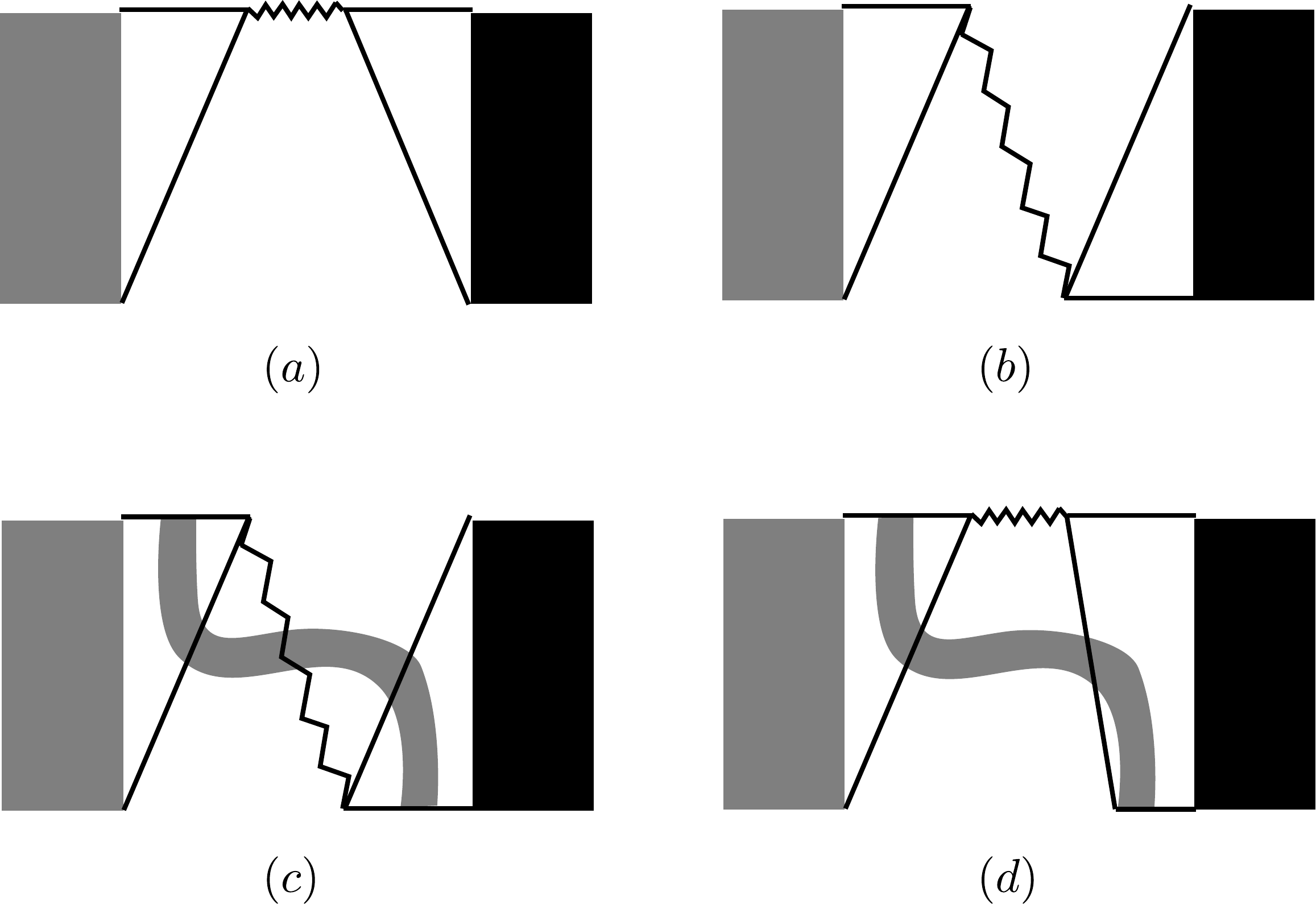}
\end{center}
\caption{Ladder insertions at $\mathcal{O}(e^2)$, in addition to the ones shown in Figure \ref{ladder}, that do not change the growth exponent, $\lambda_L$. The diagrams (a) and (b) have partner diagrams generated by reflection about the horizontal axis. The diagrams (c) and (d) have two partners each, from reflection about the horizontal and vertical axes. Other diagrams (not shown) similar to (c) and (d) with the internal resummed disorder lines terminating on the same time fold instead of opposite time folds vanish due to integrations over Green's functions with poles on the same side of the real axis.}
\label{snakes}
\end{figure}

At $\mathcal{O}(e^2)$ we have to also consider the diagrams shown in Figure~\ref{snakes}. In the diagrams given by Figure~\ref{snakes}(a), (b), the internal interaction line carries only the external frequency and momentum. We assume that the Coulomb interaction actually has a long static screening length $\xi \gg l$, where $l=v_F\tau$ is the disorder mean free path, and that we are probing scrambling at length scales $x\gg \xi$.

The interaction line in Figure~\ref{snakes}a is given by
\beq
V^A_\xi(\omega,\q) = \lim_{q\rightarrow0}\frac{4\pi e^2}{q^2+\xi^{-2}}\frac{i\omega+Dq^2}{i\omega+D\left(K^2\frac{q^2}{q^2+\xi^{-2}}+q^2\right)} = \frac{4\pi e^2}{\xi^{-2}},
\eeq 
The insertion in Figure~\ref{snakes}(a) in the limit of small external frequency and momentum $(\omega,\q)$ is then given by
\begin{align}
&Nig(0)^2\frac{\pi e^2}{\tau^2\xi^{-2}}\int\frac{d\epsilon_1}{2\pi}\frac{d\epsilon_2}{2\pi}\frac{1}{\cosh(\beta k_0/2)\cosh(\beta k_0^{\prime\prime}/2)}\frac{1}{\epsilon_1^2+1/(4\tau^2)}\frac{1}{\epsilon_1+i/(2\tau)}\frac{1}{\epsilon_2^2+1/(4\tau^2)}\frac{1}{\epsilon_2+i/(2\tau)} \nn
&=ig(0)^2\tau^2\frac{\pi e^2}{\xi^{-2}}\frac{1}{\cosh(\beta k_0/2)\cosh(\beta k_0^{\prime\prime}/2)}. 
\end{align}
The factor of $i$ comes from $(-i)^3$ from the three advanced Green's functions, and the additional factor of $N$ arises because the flavor indices on the left and the right sides of the diagram are decoupled. The partner insertion obtained by reflection about the horizontal axis is the complex conjugate of this, so their sum vanishes. For the insertion in Figure~\ref{snakes}(b), the internal Wightman line is given by
\beq
V^W_\xi(\omega,\q) = \lim_{q\rightarrow0}\frac{4\pi e^2}{q^2+\xi^{-2}}\frac{\omega}{\sinh\left(\frac{\beta\omega}{2}\right)}\frac{DK^2\frac{q^2}{q^2+\xi^{-2}}}{\omega^2+D^2\left(K^2\frac{q^2}{q^2+\xi^{-2}}+q^2\right)^2} = 0,
\eeq 
so this diagram is not important. In Figure~\ref{snakes} (c), the internal Wightman line carries the external frequency $\omega\lesssim \lambda_L\ll T$, so it can be approximated by $\frac{4\pi e^2}{DK^2p^2}T$, where $p$ is an internal momentum. However, in this case, once again, the same small momentum $p$ does not appear in both the interaction and the internal $L$ or $L_c$, so this diagram ends up being less singular and scales as subleading powers of $T$ starting at $T^2$. For Figure~\ref{snakes} (d), the internal interaction line is just $-i\frac{4\pi e^2}{K^2}$ and we get for the insertion, after appropriately shifting momenta, for both the internal $L$ and internal $L_c$ cases
\begin{align}
&\frac{Nig(0)\pi e^2}{\tau^4K^2}\int\frac{d\epsilon_1}{2\pi}\frac{d\epsilon_2}{2\pi}\frac{d^3\k}{(2\pi)^3}\frac{dk_0^{\prime\prime}}{2\pi}\frac{1}{\cosh(\beta k_0/2)\cosh(\beta k_0^{\prime\prime}/2)}\frac{1}{\left(\epsilon_1^2+\frac{1}{4\tau^2}\right)^2}\frac{1}{\epsilon_1+\frac{i}{2\tau}}\frac{1}{\epsilon_2^2+\frac{1}{4\tau^2}}\frac{1}{\epsilon_2+\frac{i}{2\tau}}\frac{1}{Dk^2-i(k_0-k_0^{\prime\prime})} \nn
&=-\frac{3Nig(0)\pi e^2\tau^2}{K^2}\int\frac{d^3\k}{(2\pi)^3}\frac{dk_0^{\prime\prime}}{2\pi}\frac{1}{\cosh(\beta k_0/2)\cosh(\beta k_0^{\prime\prime}/2)}\frac{1}{Dk^2-i(k_0-k_0^{\prime\prime})}.
\end{align}
Reflecting this insertion about the horizontal axis produces its complex conjugate, and reflection about the vertical axis effectively interchanges $k_0,k_0^{\prime\prime}$. The four contributions then sum to zero.

\subsection{Short-range interactions}

Based on the analysis of Section~\ref{QC}, we see that the Lyapunov exponent is simply given by
\begin{align}
&\lambda_L^{(1)} = -2g(0)\tau^2\int\frac{d^d\k}{(2\pi)^d}\int_{-\infty}^{\infty}\frac{dk_0}{2\pi}\frac{\mathrm{Im}[V^R(k_0,\k)]}{\sinh(\beta k_0/2)}\mathrm{Re}[L(k_0,\k)] + 4g(0)\tau^2\int\frac{d^d\k}{(2\pi)^d}\int_{-\infty}^\infty\frac{dk_0}{2\pi}\frac{\mathrm{Im}[V^R(k_0,\k)]}{\sinh(\beta k_0)}L(k_0,\k) \nn
&= -2g(0)\tau^2\int\frac{d^d\k}{(2\pi)^d}\int_{-\infty}^{\infty}\frac{dk_0}{2\pi}\frac{\mathrm{Im}[V^R(k_0,\k)]}{\sinh(\beta k_0/2)}L(k_0,\k) + 4g(0)\tau^2\int\frac{d^d\k}{(2\pi)^d}\int_{-\infty}^\infty\frac{dk_0}{2\pi}\frac{\mathrm{Im}[V^R(k_0,\k)]}{\sinh(\beta k_0)}L(k_0,\k),
\label{lambda}
\end{align}
as $\mathrm{Im}[V^R(k_0,\k)]$ and $\mathrm{Im}[L(k_0,\k)]$ are both odd functions of $k_0$ for the interactions we consider. Since $|1/\sinh(\beta k_0/2)|>|2/\sinh(\beta k_0)|$, the first term of the above (coming from the ladder sum of Figure~\ref{ladder}) always dominates the second (coming from the self-energy corrections), and the exponent is thus always positive if $\mathrm{sgn}(\mathrm{Im}[V^R(k_0,\k)])=-\mathrm{sgn}(k_0)$. For a short-range interaction that does not vanish as $q\rightarrow0$ (we take a contact interaction for which $V_{bs}^R(\q)=V_0$), screening by the diffuson produces
\beq
V^R_s(\omega,\q) = V_0\frac{-i\omega+Dq^2}{-i\omega+D^\prime q^2},~~\mathrm{Im}[V^R_s(\omega,\q)]=V_0\frac{\omega(D-D^\prime)q^2}{\omega^2+D^{\prime2}q^4},~~D^{\prime} = D\left(1+\frac{dn}{d\mu}V_0\right)>D.
\eeq 
Inserting this into Eq.~(\ref{lambda}), we see that all the integrals converge, and that $\lambda_L^{(1)}\sim + V_0^2T^{3/2}$ for $d=3$. Thus, short-range interactions behave qualitatively in the same way as Coulomb interactions from the point of view of scrambling, consistent with previous work on the inelastic scattering rate~\cite{Lee1985}.    
\end{widetext}

\section{Two spatial dimensions}
\label{2d}

In two spatial dimensions, the diffuson-screened Coulomb interaction is~\cite{Lee1985}
\beq
V_2^R(\omega,\q) = \frac{2\pi e^2}{q}\frac{-i\omega+Dq^2}{-i\omega+DK_2q+Dq^2},~~K_2=2\pi e^2\frac{dn}{d\mu}.
\eeq
We probe scrambling at length scales $\x$ much larger than the mean free path $l$ and the screening length $K_2^{-1}$ but smaller than the eventual localization~\cite{PWA58} length $l~e^{k_Fl}$ of the electron wavefunctions~\cite{Lee1985} (The light-cone like growth of $f(t,\x)$ will be arrested beyond this localization length, i.e. the operator-radius $\R(t)$ is bounded by this length). Then, the same approximations and lines of reasoning we used in three dimensions also work in two dimensions, and the Lyapunov exponent is still given by Eq.~(\ref{lambda}) with $d=2$. Inserting this dynamically screened Coulomb interaction, we obtain the leading contribution
\begin{align}
\lambda_{L2}^{(1)} & =  \frac{e^2}{2DK_2}\int_0^\infty dk_0\left(\frac{1}{\sinh(\beta k_0/2)}-\frac{2}{\sinh(\beta k_0)}\right) \nn & = \frac{e^2 T}{D K_2}\ln2 = \frac{T}{2\pi D(dn/d\mu)}\ln2 \nn & \approx\frac{e^2R_\square}{h}\frac{k_B T}{\hbar}\ln 2,
\label{lambda2}
\end{align}
where $R_\square=1/(e^2D(dn/d\mu))$ is the sheet resistivity~\cite{Efros2012} and we restored factors of $k_B$ and $\hbar$. This cannot saturate the universal bound $\lambda_L\le 2\pi k_B T/\hbar$ unless the effective coupling $e^2R_\square/h$ becomes large, which also determines crossover or transition to an insulating state. According to experimental results reported in Ref.~\onlinecite{Kravchenko1996} and theory discussed in Ref.~\onlinecite{Castellani1998}, the density-tuned metal-insulator crossover/transition occurs at around $R_\square \approx 3h/e^2$, which is smaller than the value required to saturate the bound by about a factor of $3$. This indicates that the metallic state has a Lyapunov exponent numerically, but not parametrically, smaller than the bound. 

From Eq. (\ref{lambda2}) above, we see that it contains the difference of two terms. The term being subtracted is the decay rate of electrons in exact eigenstates of the disorder potential~\cite{Abrahams1981,Fukuyama1983}, whereas the term being added gives the rate at which chaos spreads, i.e. how electrons would be infected within an epidemic picture (See Figure~\ref{introfig}) if there were no electron `deaths'. Both these terms individually contain a logarithmic infrared divergence, which cancels when their difference is taken. The logarithmic divergence in the exact eigenstate decay rate was removed in a self-consistent computation~\cite{Fukuyama1984}, by using the rate itself as an infrared energy cutoff, but this is not required here. For the exact eigenstate decay rate, the self-consistent computation provides instead a regularized logarithmic factor of $\ln(\pi D (dn/d\mu))$~\cite{Fukuyama1984,Efros2012,Lee1985}, which doesn't appear in the Lyapunov exponent.

Let us comment now on why the logarithmic divergence cancels out in the expression for the Lyapunov exponent but appears in the exact eigenstate decay rate. It arises from an infrared divergence in the collision integral in Eq.~(\ref{sig}) when the energy transfer in a collision approaches zero. At zero energy transfer, the interaction of the electrons with another particle-hole excitation (or equivalently the boson representing the Coulomb interaction) is like the electrons scattering off a random static potential.  Each instance of such a scattering event can be described by a quadratic integrable Hamiltonian, and is hence incapable of producing chaos. However, this process still leads to decoherence of the individual electron wavepackets and hence contributes to the decay rate. A similar cancellation between singular pieces of self-energy and ladder contributions coming from zero energy transfer collisions was first pointed out by two of us in the computation of the Lyapunov exponent of a Fermi surface coupled to a gapless fluctuating gauge field in Ref.~\onlinecite{Patel2017}. For the short-range interactions considered in the previous section, the logarithmic factor still cancels in the Lyapunov exponent, and we obtain $\lambda_{L2}^{(1)}\sim +V_0^2 T$.       

\section{Discussion}
We have studied the spread of many-body quantum chaos due to electron-electron interactions in diffusive metals. We find that chaos spreads {\it ballistically}, even though quasiparticles are transported {\it diffusively}. This is because the spread of chaos is linked only to the propagation of quantum information about inelastic collisions of quasiparticles, which does not require the transport of quasiparticles themselves. In three dimensions, we found that the Lyapunov exponent scales as the inelastic scattering rate of quasiparticles, whereas in two dimensions the inelastic scattering rate is larger than the Lyapunov exponent by a logarithmic factor arising from `classical' collisions that do not involve quantum fluctuations. In $d$ spatial dimensions, we find $\lambda_L\sim T^{d/2}$, which leads to $v_B\sim T^{d/4}$. Comparing the form of the butterfly velocity to a scaling form $v_B \sim T^{1-1/z}$, where $z$ is the dynamical exponent, we find that our result is qualitatively similar to that of a critical system with $z>1$. While our computations in $d=2$ and 3 were carried out with the $1/r$ Coulomb interaction, we expect similar results to hold in $d=2$ for the $\ln r$ Coulomb interaction.

Remarkably, we find the above ballistic growth of operators even though the Coulomb interaction is long ranged and no microscopic Lieb-Robinson bound exists. This result is a particularly striking example of the idea that the butterfly velocity can function like a low energy Lieb-Robinson velocity \cite{BSLR}. It raises the question of what other long range models might be harboring an emergent ballistic growth of operators at low energy.

We note the recent experimental measurement by Kapitulnik {\it et. al.\/} of local thermal diffusivity using an optical method~\cite{Zhang2016}. It would be interesting to measure the local heat diffusion constant in an interacting diffusive metal using this method. The heat diffusion constant is given by the ratio of thermal conductivity and specific heat; at low enough temperatures, in a regime where both of these quantities are dominated by the electronic contribution, it would be interesting to compare the measured diffusion constant to the known quasiparticle diffusion constant $D$ that appears to be relevant to quantum chaos. While in the non-interacting case one expects the thermal diffusivity to be equal to $D$, significant deviations may arise due to interactions, especially in two dimensions~\cite{Catelani2005}.

In this work we only focused on disorder averaged correlation functions in the diffusive, ergodic phase. However, one could also ask about rare-region effects \cite{TV10,KA15}. For example, can rare `localized' regions in the ergodic phase impede the spread of chaos? How does the different inelastic scattering rate in these regions~\cite{FA80} affect the Lyapunov exponent? Alternatively could there be rare-regions, with very little disorder, that lead to an even faster butterfly velocity? In dimensions greater than one, the effect of such rare-regions are expected to be significantly suppressed, but we leave a detailed study for future work. Finally, it would also be interesting to study the growth of entanglement in an interacting diffusive metal, and compare it to the spread of chaos. We also leave this question for future study.

\section*{Acknowledgements}
AAP and SS acknowledge support by the NSF under Grant DMR-1360789. SS and BGS acknowledge support from a MURI grant W911NF-14-1-0003 from ARO. BGS is also supported by the Simons Foundation as part of the It From Qubit collaboration and through a Simons Investigator Award to Senthil Todadri. DC is supported by a postdoctoral fellowship from the Gordon and Betty Moore Foundation, under the EPiQS initiative, Grant GBMF-4303, at MIT. Research at Perimeter Institute is supported by the Government of Canada through Industry Canada and by the Province of Ontario through the Ministry of Research and Innovation. SS also acknowledges support from Cenovus Energy at Perimeter Institute.

\appendix
\section{Outline of Feynman rules for the complex-time contour}
\label{FR}

In this appendix we briefly outline the Feynman rules on the complex-time contour work that are used to compute Eq.~(\ref{oto}). A detailed derivation of Feynman rules for such scenarios has been presented earlier in Refs.~\cite{Stanford2016,DCBS17}. We split the Hamiltonian $H$ into three pieces corresponding to the clean, non-interacting system, the disordered potential, and the interaction term
\beq
H = H_0 + H_{\mathrm{int}} \equiv H_{\mathrm{free}}^{\mathrm{clean}} + H_{\mathrm{free}}^{\mathrm{dis}} + H_{\mathrm{int}}
\eeq

For $H_{\mathrm{free}}^{\mathrm{clean}}$, Eq.~(\ref{oto}) simply factorizes by Wick's theorem into a product of a retarded Green's function and an advanced Green's function. When disorder is included, we have
\begin{align}
\psi^{\mathrm{dis}}_{\mathrm{free}}(t,\x) = & (\mathcal{T}e^{-i\int_0^t dt H_{\mathrm{free}}^{\mathrm{dis}}[t,\psi^{\mathrm{clean}}_{\mathrm{free}}]})\psi^{\mathrm{clean}}_{\mathrm{free}}(t,\x) \nn & \times (\mathcal{T}e^{-i\int_0^t dt H_{\mathrm{free}}^{\mathrm{dis}}[t,\psi^{\mathrm{clean}}_{\mathrm{free}}]})^\dagger,
\end{align}
where $\mathcal{T}$ denotes time-ordering. The exponentials containing $H_{\mathrm{free}}^{\mathrm{dis}}$ may now be expanded, this produces corrections to Eq.~(\ref{oto}) with $H=H_{\mathrm{free}}^{\mathrm{clean}}$ that can be contracted by Wick's theorem and the disorder average Eq.~(\ref{disavg}). Since the disorder is time-independent, this produces to lowest order the disorder self-energy corrections to the Green's functions (Figure~\ref{basic}a), and also the disorder ladder corrections in Figure~\ref{bs0}a. These corrections can then be resummed to obtain the non-interacting $f(t,\x)$ as shown in Appendix~\ref{NI}.

With the inclusion of interactions, we use
\begin{align}
\psi(t,\x) & = (\mathcal{T}e^{-i\int_0^t dt( H_{\mathrm{free}}^{\mathrm{dis}}[t,\psi^{\mathrm{clean}}_{\mathrm{free}}]+H_{\mathrm{int}}[t,\psi^{\mathrm{clean}}_{\mathrm{free}}])})\psi^{\mathrm{clean}}_{\mathrm{free}}(t,\x) \nn & \times (\mathcal{T}e^{-i\int_0^t dt( H_{\mathrm{free}}^{\mathrm{dis}}[t,\psi^{\mathrm{clean}}_{\mathrm{free}}]+H_{\mathrm{int}}[t,\psi^{\mathrm{clean}}_{\mathrm{free}}])})^\dagger.
\end{align}
It is helpful to consider for the purposes of this illustration the decoupling the four-Fermion interactions using a bosonic field $\varphi(\omega,\k)$ with a propagator given by the unscreened Coulomb interaction $V_b(\k)$. The perturbative expansion now generates the corrections shown in Figure~\ref{se} that involve the usual correction to the Green's functions due to interactions, along with a new set of corrections that involve the contraction of the boson field across the $e^{-\beta H/2}$ thermal factors of Eq.~(\ref{oto}),
\beq
V^W(\omega,\k) \equiv \mathrm{Tr}[e^{-\beta H/2}\varphi(\omega,\k)e^{-\beta H/2}\varphi(-\omega,-\k)].
\eeq
The expression for this Wightman propagator $V^W$ is provided in Eq.~(\ref{vwight}), and its relation to the spectral function is derived in detail in Refs.~\cite{Patel2017,DCBS17}. These new corrections generate the diagrams shown in Figure~\ref{ladder}. Note that interaction corrections to the $e^{-\beta H/2}$ thermal factors in Eq.~(\ref{oto}) correspond to the dressing of the Wightman propagators, which we take into account since we use the dynamically screened Coulomb interaction for $V^W$ in Eq.~(\ref{vwight}).

\begin{widetext}
\section{Absence of chaos in the non-interacting disordered metal}
\label{NI}

In this appendix we derive the expression for $f(t,\x)$ in the non-interacting scenario. We have (see Eq. \eqref{oto} and Figure~\ref{bs0}),
\begin{align}
&f(\omega,\q) = \int \frac{d^d\k dk_0}{(2\pi)^{d+1}}G^R_0(k_0+\omega,\k+\q)G^A_0(k_0,\k) + \nn 
&\int \frac{d^d\k_1d^d\k_2}{(2\pi)^{2d}}\frac{dk_0}{2\pi}G^R_0(k_0+\omega,\k_1+\q)G^A_0(k_0,\k_1)G^R_0(k_0+\omega,\k_2+\q)G^A_0(k_0,\k_2)L(\omega,\q).
\end{align}
The diffuson rung $L(\omega,\q)$ is given by the following resummation of disorder rungs:
\begin{align}
&L(\omega,\q)=U_0^2+U_0^2\int\frac{d^d\k}{(2\pi)^d}G^R_0(k_0+\omega,\k+\q)G^A_0(k_0,\k)L(\omega,\q) \nn
&=U_0^2 + U_0^2\int \frac{d^d\k}{(2\pi)^d} \frac{1}{\frac{k^2}{2m}-\mu-k_0+\frac{i}{2\tau}}\frac{1}{\frac{(k+q)^2}{2m}-\mu-k_0-\omega-\frac{i}{2\tau}}L(\omega,\q) \nn 
&\approx U_0^2 + U_0^2\int \frac{d^d\k}{(2\pi)^d} \frac{1}{\frac{k^2}{2m}-\mu-k_0+\frac{i}{2\tau}}\frac{1}{\frac{k^2}{2m}-\mu-k_0-\frac{i}{2\tau}}\left(1+\frac{\omega}{\frac{k^2}{2m}-\mu-k_0-\frac{i}{2\tau}}+\left(\frac{\k\cdot \q/m}{\frac{k^2}{2m}-\mu-k_0-\frac{i}{2\tau}}\right)^2\right) L(\omega,\q) \nn
&\approx U_0^2 + U_0^2g(0)\int \frac{d\epsilon}{2\pi} \frac{1}{(\epsilon-k_0)^2+\frac{1}{4\tau^2}}\left(1+\frac{\omega}{\epsilon-k_0-\frac{i}{2\tau}}+\frac{q^2v_F^2/d}{\left(\epsilon-k_0-\frac{i}{2\tau}\right)^2}\right) L(\omega,\q). \\
&L(\omega,\q) =\frac{1}{g(0)\tau^2(-i\omega+Dq^2)}. 
\end{align}
where $D=v_F^2\tau/d$, and in the intermediate steps, we expanded in small $q$ assuming that the largest contributions to the integrals come from the regions with $k\sim k_F = m v_F\gg q$, and that $\mu\gg\tau^{-1}\gg|\omega|$. We assumed that $L(\omega,\q)$ does not depend on any other combinations of momenta and frequencies passing through it apart from ($\omega,\q$), which turns out to be self-consistent. Each disorder rung is multipled by a factor of $-i^2=1$, where the $i$'s come from the real-time electron-disorder vertices. We thus see that $f(t,\x) \sim f_0(t,\x) + f_1(t,\x) e^{-x^2/(4Dt)}$, where $f_0$ decays rapidly in time at a rate given by $\tau^{-1}$ and $f_1$ is a slowly varying function of space and time. Henceforth we ignore $f_0$ as we are interested in long times $t\gg\tau$ and set $f_1$ to 1. Since there is no exponential growth in $f(t,\x)$ we conclude that the non-interacting disordered metal does not have many body quantum chaos.

\end{widetext}

\bibliography{dirt}

\end{document}